\documentstyle[preprint,aps]{revtex} 
\begin{document}
%\draft command makes pacs numbers print
\draft
\widetext
%\narrowtext
\title{ The extended Hubbard model applied to study the pressure effects in 
high temperature superconductors     }
\author{E. V.L. de Mello\cite{email} and C. Acha} 
\address{Centre National de la Recherche Scientifique,\\ Centre 
de Recherche sur les Tr\`{e}s Basses Temp\'{e}ratures\\
        Boite Postale 166, 38042 Grenoble-Cedex, France. \\}
\date{\today}
\maketitle
\begin{abstract}
 We  make use of a  BCS type approach  based on the extended Hubbard
Hamiltonian to study the superconductor transition  
and  to give a microscopic interpretation
of the pressure effects on $T_c$ in high temperature superconductors. 
This novel method suggests that the applied pressure
causes an increase of the superconducting gap and this effect is explored
in order to explain the variations of $T_c$. Our  approach is therefore
beyond the scope of previously phenomenological models which basically
postulate a  pressure-induced charge transfer and an intrinsic term
linear on the pressure. We obtain a microscopic interpretation of this
intrinsic term and a general expansion of $T_c$ in terms of the pressure.
To demostrate the
efficiency  of the method we apply  it to
the experimental data of the Hg-based superconductors.
\end{abstract}
\pacs{71.10.Fd,74.62.Fj,74.25.Dw}
%\narrowtext
%\vskip[2pc]

\section{Introduction}

After the discovey of high temperature superconductors (HTSs) it was
realized that the critical temperature ($T_c$) could be substantially
enhanced  by applying high pressures. This very interesting effect
attracted a lot of attention as it is summarized by 
several  review articles which have been written on the 
subject\cite{Griessen,Taka,Schill}. The motivation for all
this effort is not only the quest for higher temperatures but also to 
understand which are the parameters which optimise $T_c$ besides the 
possibility of follow the structural changes induced by  pressure by
means of x-rays and neutron difraction techniques. It is hoped that
these investigations lead  to the preparation of new materials by 
chemical substitution as well as some better insights on these
highly complex systems.

One of the effects of the  pressure which is generally 
accepted and well documented in certain materials\cite{Griessen,Taka,Schill}
is an increase of the carrier concentration on the $CuO_2$
planes transferred from the reservoir layers. Such pressure induced
charge transfer (PICT) has been confirmed by Hall effect and
thermoeletric power measurements in several compounds\cite{Taka}.
Therefore this effect combined with an assumption of an  intrinsic 
variation  of $T_c$
(linear on the pressure) independent of the charge 
transfer was largely explored to account for the 
quantitative relation between $T_c$ and the pressure P and gave
origin to many models\cite{Griessen2,Almasan,Neumeier,Gupta,Mori}. 
Some of 
these models also invoked that several HTSs have a $T_{c}$ versus
carrier density  $n$ (per $CuO_2$) diagram which satisfies 
a phenomenological
universal parabolic curve, i.e., $T_{c}=T_c^{max}[1-\eta(n-n_{op})^{2}]$
where $n_{op}$ is the optimum $n$. Following along these lines we
can write $n(P)=n+\Delta n(P)$,  $T_{c}=T_{c}(n,P)$ and 
derive an expansion in powers of P, namely 
\begin{equation}
 T_{c}(n,P)=T_{c}(n,0)+\alpha_1(n)P+\alpha_2(n)P^2
\label{exp1}
\end{equation}
where the coefficients are, 
\begin{mathletters}
\label{allequations}
\begin{equation}
\alpha_1(n)=\partial T_{c}(n)/\partial{P}
-2\eta T_c^{max}(n-n_{op})\partial{n}/\partial{P} 
\label{alpha}
\end{equation}
and
\begin{equation}
 \alpha_2(n)= -\eta T_c^{max}(\partial{n}/\partial{P})^2.
\label{alphab}
\end{equation}
\end{mathletters}
Where the first term in $\alpha_1$ is known as the intrinsic term and 
it was estimated to 
be constant\cite{Griessen2,Almasan,Neumeier,Gupta,Mori,Griessen3}. 
This approach
was largely used  in describing the data in the vicinity of 
$n_{op}$\cite{Taka,Neumeier}
but fails to describe more recently measurements on different compounds
with a large variation
on the initial $n$ values from underdoped to overdoped regime\cite{Taka,Cao}.
Futhermore this phenomenological method  does not provide  any
physical insight on the  origin of the (charge transfer independent)
intrinsic term. 

\section{Method and Approach}

We propose in this work some new ideas which are general enough to
be applied to any family of compounds and which provide some microscopic
interpretation on  the effects of 
pressure. We use as starting point a recently introduced approach\cite{Mello}
based on a BCS type mean field analysis  which uses the 
extended Hubbard Hamiltonian (t-U-V)
on a square lattice (of lattice parameter a) 

\begin{equation}
H=-\sum_{\langle ij \rangle , \sigma}t(c^{\dagger}_{i \sigma} c_{j \sigma}
 + h.c.)+U \sum_{i} n_{i \downarrow}n_{i \uparrow}-V \sum_{\langle ij
 \rangle}n_{i}n_{j} .
\label{Hamiltonian}
\end{equation}

For the sake of completeness we briefly outline
the method. A gap equation at zero temperature is derived which has
the same form as in  the usual BCS theory, i.e.,

\begin{equation}
\Delta_{\vec k}=-\sum_{\vec l}V_{\vec k\vec l}\frac{\Delta_{\vec l}}
{2\left(\xi^2_{\vec l}+\Delta^2_{\vec l}\right)^{1/2}} .
\label{gap}
\end{equation}
Where $\xi_{\vec k}=-4t(cos(k_{x}a)+cos(k_{y}a))-\mu$ , 
$V_{\vec k\vec l}$ is
the Fourier transform of the potential of Eq.\ (\ref{Hamiltonian}), 
which is approximately given by
\begin{equation}
V_{\vec k\vec l}\approx U-4V\left(\cos(k_x-l_x)a+
\cos(k_y-l_y)a\right) .
\label{kpot}
\end{equation}                 

As in the  BCS  mean field method\cite{Mello} the gap  
has the same functional form of the potential, namely, $\Delta_{\vec k}=
\Delta_0 (cos(k_{x}a)\pm cos(k_{y}a))/2$, where the plus sign is for the 
s-wave and the minus sign is for d-wave channel. The chemical potential $\mu$
must be calculated self-consistently but as it concerns 
the superconducting phase 
boundary it suffices to approximate it by the value of the maximum energy
(concentration dependent) in a tight-binding  band. One then
derives a finite temperature analog of Eq.\ (\ref{gap}) and in order to 
determine the phase boundary we take  $T=T_c$ at the limit
where $\Delta_0=0$ is applied. The zero temperature and 
the finite temperature with
$T=T_c$ gap equations are solved numerically and one 
matches $T_c$ with $\Delta_0$
for a given carrier concentration $n$. Thus one obtains a phase diagram
$T_{c}\times n$ for a given $\Delta_0$ which was found to reproduce well
the experimental  phase diagrams for the HTSs when the 
position of the attractive potential V
was changed from the original nearest neighbor position and 
became an adjustable parameter\cite{Mello}. The ratio of U/V 
is not relevant  to determine the phase diagram  boundary  as long
as $U\gg V$ but, on the other hand, it determines the value of
$\Delta_0$. The exact calculations of $\Delta_0$ in terms of U  and V
are not easy to be performed in a many-body system 
and thus  it becomes a second  parameter to be determined by 
comparison with the experimental $T_c \times n$ phase diagrams. 
It was shown\cite{Mello}  
that the chosen values for $\Delta_0 $ that reproduce the phase diagrams
of the Lanthanum and Yttrium family of compounds are also
in excelent agreement with the gap measurements taken from 
tunneling experiments and the  specific heat.
As concerns the  ratio of the positions of V used for these   
families, it was also shown a posteriori that their values
matches their  ratio of the coherence length
(as discussed in Ref.\cite{Mello}, these are strongly coupled systems
and the bound states are confined which is not the case of weakly
coupled systems) thus providing a possible physical explanation
for this quantity and why they are so different for the La and Y
families of compounds.

In connection with the above discussion we are led to propose that
the effects of pressure are two-fold: (i)- The well accepted PICT;
(ii)- The relation $2\Delta_{0}=\gamma k_{B}T_c^{max}$ ($\gamma=3.5$
for weakly BCS and $\gamma\approx4.3$ for the method
mentioned above) suggests that if $T_c^{max}$ is a linear
function of the pressure P (as assumed for the behavior of the
intrinsic term\cite{Almasan,Neumeier,Gupta}) 
than  $\Delta_0$ must also be a linear function of P. As concerns 
the  Hamiltonian of Eq.(3) this
is equivalent to say that the structural modifications due to the applied
pressure induce  a variation on the attractive potential V (which is the
most influencial parameter on the value of $\Delta_0$). In fact the real effect
of the structural changes can only be estimate by electronic band
calculations\cite{Taka,Schill,Novikov,Gupta2},  but they are 
not  adequate to be performed in the context of
the strong correlations of the t-U-V Hamiltonian.
 
Thus the PICT (i) implies that $n(P)=n+\Delta n(P)$ and the assumption
of a pressure dependent gap (ii) implies $\Delta_0(P)=\Delta_0+
\Delta \Delta_0(P)$ which
can be simple written as $T_{c}=T_{c}(n(P),\Delta_0(P))$.
On Fig.1 we plot two curves calculated with two differents values
of $\Delta_0$ to study
how $T_c(n)$ change if the pressure induces a change in  $\Delta_0$.
Therefore to estimate $T_c$ of a compound with an initial 
given value of $n$ and
under a given pressure P,  we  perform an expansion of 
$T_c(n)$ in terms of P. With the assumption of linear variation of $n$ and
$T_c^{max}$ (or $\Delta_0$) on the pressure,  we obtain only 
terms up to third order, that is,
\begin{mathletters}
\label{allequations}
\begin{equation}
T_c(n,P)=\sum_{Z=0}^3 \alpha_Z P^Z
\label{expP}
\end{equation}
with
\begin{equation}
\alpha_Z=({\partial \over \partial \Delta_{0}}{\partial \Delta_0 \over
 \partial P} + {\partial \over \partial n}{ \partial n \over \partial P})^Z
 T_c(n(P),\Delta_0(P))
\label{coeff}
\end{equation} 
\end{mathletters}

Furthermore one can derive simple analytical expressions for
each coefficient in an approximate way,  using the universal 
parabolic fitting and with $T_{c}^{max}(P)=2\Delta_{0}(P)/\gamma$ 
which explicits the $T_c$ dependency on P (assumption ii).
This procedure gives an intrinsic term which is radically different than 
that used before\cite{Griessen2,Almasan,Neumeier,Gupta,Mori}
as well as a new third order term. 

\section{Comparison with the Experiments}

To illustrate the entire method we will apply it to the 
mercury system. The mercury family of compounds represents a 
real challenge to any theory for the following reasons: (a)- The
highest transition temperatures obtained up to date have been
measured on Hg1223 at 25-30 GPa\cite{Chu,MNR,Gao,Jover} reaching 
values up to 164K. (b)-The various pressure  data for the underdoped and
overdoped compounds of $HgBa_2CuO_{4-\delta}$ (Hg1201) could
not be interpreted\cite{Cao} by the  models described in the
introduction. (c)- The
largest pressure effect on $T_c$ with  a change of 
almost 50K over a span of 20 GPa has been
recently  measured by one of us\cite{Acha} in the Hg2212.

As a preeliminary step and also in order to determine the initial
parameters,   we need to study the $T_c\times n$ phase diagram 
at zero pressure.
Thus  we  perform a calculation using the method of 
de Mello\cite{Mello} to obtain  a  Hg1201 phase diagram which
is in agreement with the experimental results\cite{Cao}. As
discussed above this method of calculation involves two parameters;
our best result is obtained with $\Delta_0=210K$ 
and the position of the attractive potential V at the $6^{th}$-neighbor.
Our results are plotted on Fig.1  and, just for comparison and 
for future use,  the phenomenological  
parabolic fitting with $\eta=50$ and $T_c^{max}=97$K as used
by Cao et al\cite{Cao}. Thus the  calculation for  the Hg family
phase diagram is our first step and it is independent of the
pressure studies which we will deal next.

To study the effects of the pressure we  also plotted
on Fig.1 our calculation for the phase diagram with 
$\Delta_0=240K$. We see that near 
$n_{op}=0.16$ the variations of $T_c$ with respect to $\Delta_0$ 
are  almost twice as those at the extremes (low $T_c$)  and since this
is the origin of the intrinsic term,  it also varies
in the same way and this behavior will be discussed below.  
It is important to notice that the two partial derivatives
that appear  in Eq.\ (\ref{coeff}) become two parameter to be
determined by comparing with the $T_c\times P$ measurements for two 
compounds with different charge density $n$. 
At low pressures only the linear terms $\alpha_1$ 
comes into play since the higher order coefficients are negligible.
It is desirable (but not crucial) to start with  $n=n_{op}$
to   determine $\partial \Delta_0 / \partial P$ since at
$n_{op}$  the charge transfer term vanishes. So with other set
of $T_c \times P$ data at another value of $n\not= n_op$, we determine 
$\partial n / \partial P$.  After these two parameters being determined
we can apply the expansion Eq.\ (\ref{expP}) to any other compound
with different value of $n$.

To illustrate the above general procedure, we will 
apply it  to the low pressure 
results of Cao et al\cite{Cao} for the Hg1201 and those of higher
pressures of Gao et al\cite{Gao}. Our purpose is to show  
that we can describe all their
results with a single choise of parameters. Thus to obtain
the value of the two partial derivatives of Eq.\ (\ref{coeff}) pertinent
to their  measurements we do the following;
we start with  the set of data taken with the compound with $n_{op}=0.16$, 
which  has a $T_c \times P$ curve at low pressures  that  is a straight
line and from which we can infer that the  linear 
term $\alpha_1=1.85$K/GPa, which at $n_{op}$,
is equal to the intrinsic term and from this we determine 
$\partial \Delta_0 / \partial P$. To determine the other
partial  derivative we study the $T_c \times P$ plot made with the sample
with $n=0.06$. We see that the low pressure slope ($\alpha_1$) can 
be taken as  $\alpha_1=2.6$K/GPa. At $n=0.06$ the intrinsic term is
half of that at $n_{op}$ according the discussion on the above
paragraph and therefore, the intrinsic term is
$0.9$K/GPa and then the  charge transfer must be  equal to $1.7$K/GPa 
since the sum of both terms is equal to $\alpha_1$. Using 
now the explicity expression
for $\alpha_1$ with the  parabolic fitting with $T_c^{max}=97$K
and $\eta =50$  we can derive   that
$\partial n/\partial P =1.8\times 10^{-3}$holes/GPa.
We notice that  this value that we obtain for $\partial n/\partial P $
is very close to others estimations  
\cite{Almasan,Gupta,Acha,Griessen3}. Thus   with this procedure
the two derivatives which comes in the calculations of
the coefficients of  Eq.\ (\ref{coeff}) are determined and
we are set now to apply Eq.\ (\ref{expP}) to any Hg compound.  
Far from $n_{op}$ the charge transfer term  dominates over the 
intrinsic one and thus the linear term $\alpha_1$ varies from the 2.6K/GPa
chosen above at n=0.06 
up to -1.0K/GPa  (at n=0.26) in the overdoped region. 
The results of our calculations  for all other compounds with different 
densities $n$  are in  excellent agreement with the low pressure
experimental data in both the underdoped and overdoped regime and 
are plotted  on Figs.2a and 2b. In Fig.3 we
show the results for  the high pressures measurements
for the three family  of mercury\cite{Gao} at $n_{op}$. 
As one can see on Fig.3 at the high pressures the quadratic 
and the cubic terms in the pressure expansion become
important (for $P>20$ GPa) and  the agreement with the data
is also remarkable. It is very interesting that the parameters
obtained above at the low pressures yielded results with the maxima
around 30 GPa which is the value of highest $T_c$ for a HTSs\cite{Gao}.
Our results can also be applied to others measurements 
on  differents families and compounds of Hg which, at $n_{op}$,
also yielded the $T_c$ maxima around 30 GPa\cite{Jover}.
For the first time a simple theory
is capable to describe successfully all this ensemble of low
and high pressure data and furthermore provides  some insights on the 
microscopic effects of the pressure.

\section{Conclusions}

Thus, we  conclude pointing out that our novel calculations (based on
a BCS type mean field with the extended Hubbard Hamiltonian) demonstrate
to be highly successful to describe the effects of the pressure with
just two simple assumptions. The PICT which is  well accepted and 
that of the pressure induced variation  of the 
superconductor gap which was introduced, 
to our knowledge, in this work. We hope that this assumption
can be confirmed in  the future by in situ experiments like 
specific heat and tunneling measurements.
As this work  came to completion, we learnt about the  work of
Angilella et al\cite{Angilella} which also uses the extended Hubbard
model within a BCS type approach.  Furthermore they estimate the 
effects of the pressure on the attractive potential V
which goes along with the lines of our assumption over $\Delta_0$.
They also obtained very good results for the effects of pressure
in the Bi2212 family, given more support to the BCS mean field
calculations with the extended Hubbard as a model for the
the physics of the charge carriers in the  HTSs. 

\section{Acknowledgments}
 
E.V.L de Mello acknowledges profitable discussions with
prof. Roman Micnas who called our attention to the work of Angilella et al
prior to its publication and  thanks CNPq of Brazil 
for a postdoctoral fellowship.
C.Acha thanks the CONICET of Argentina for a postdoctoral fellowship.

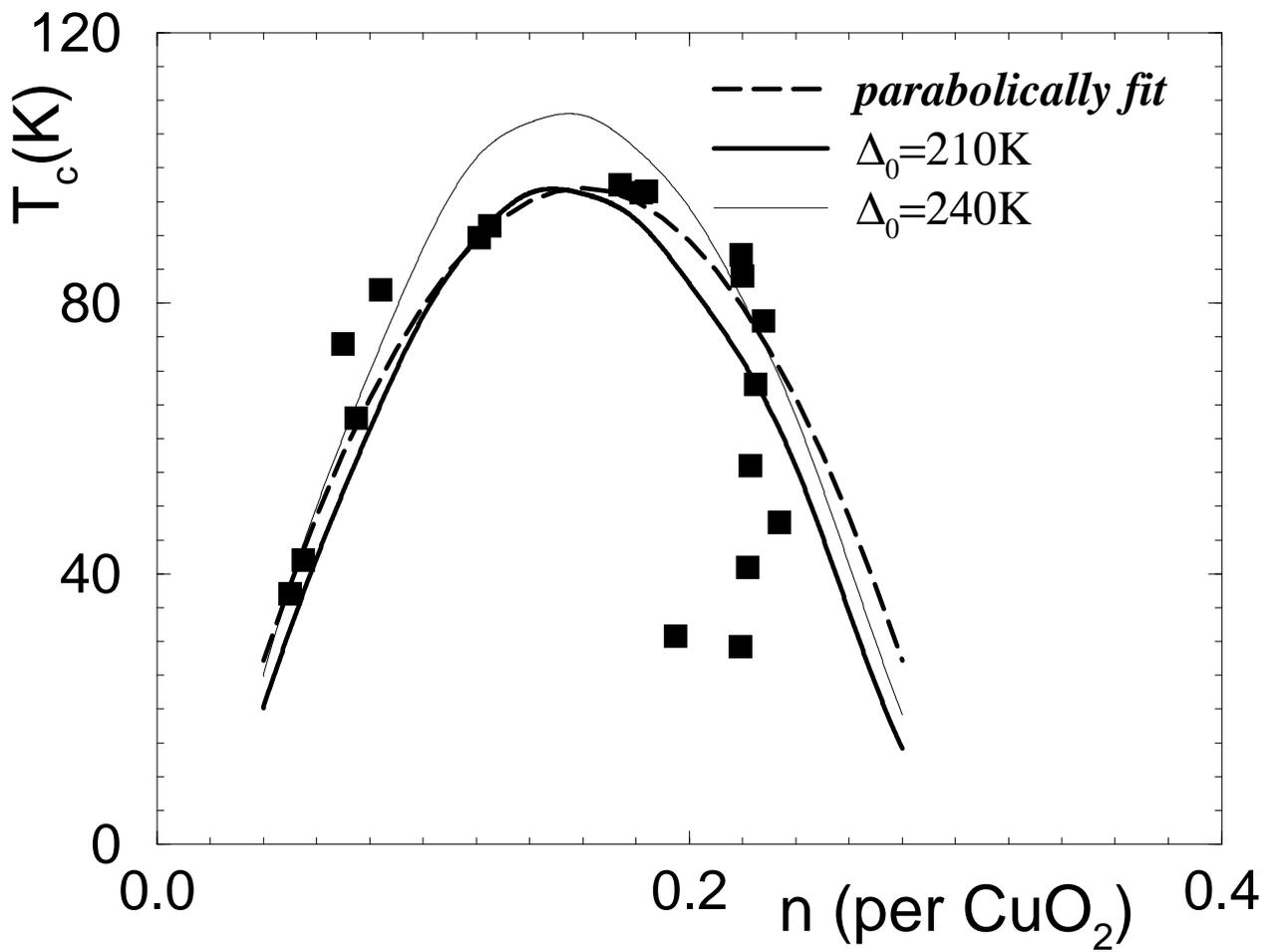
\begin{figure}
\caption{ The phase diagram for the $HgBa_2CuO_{4-\delta}$. The 
squares are the experimental points of Ref.6. Our best
fitting with $\Delta_0=210$K and the results with $\Delta_0=240$K
to illustrate the effect of changing this parameter.} 
\label{fig1}
\end{figure}

\begin{figure}
\caption{ a) The lines are our caculations for the underdoped 
region in comparison of the experimental points of Ref.10.
b)  The same  caculations for the overdoped 
region in comparison of the experimental points of Ref.10.}
\label{fig2}
\end{figure}

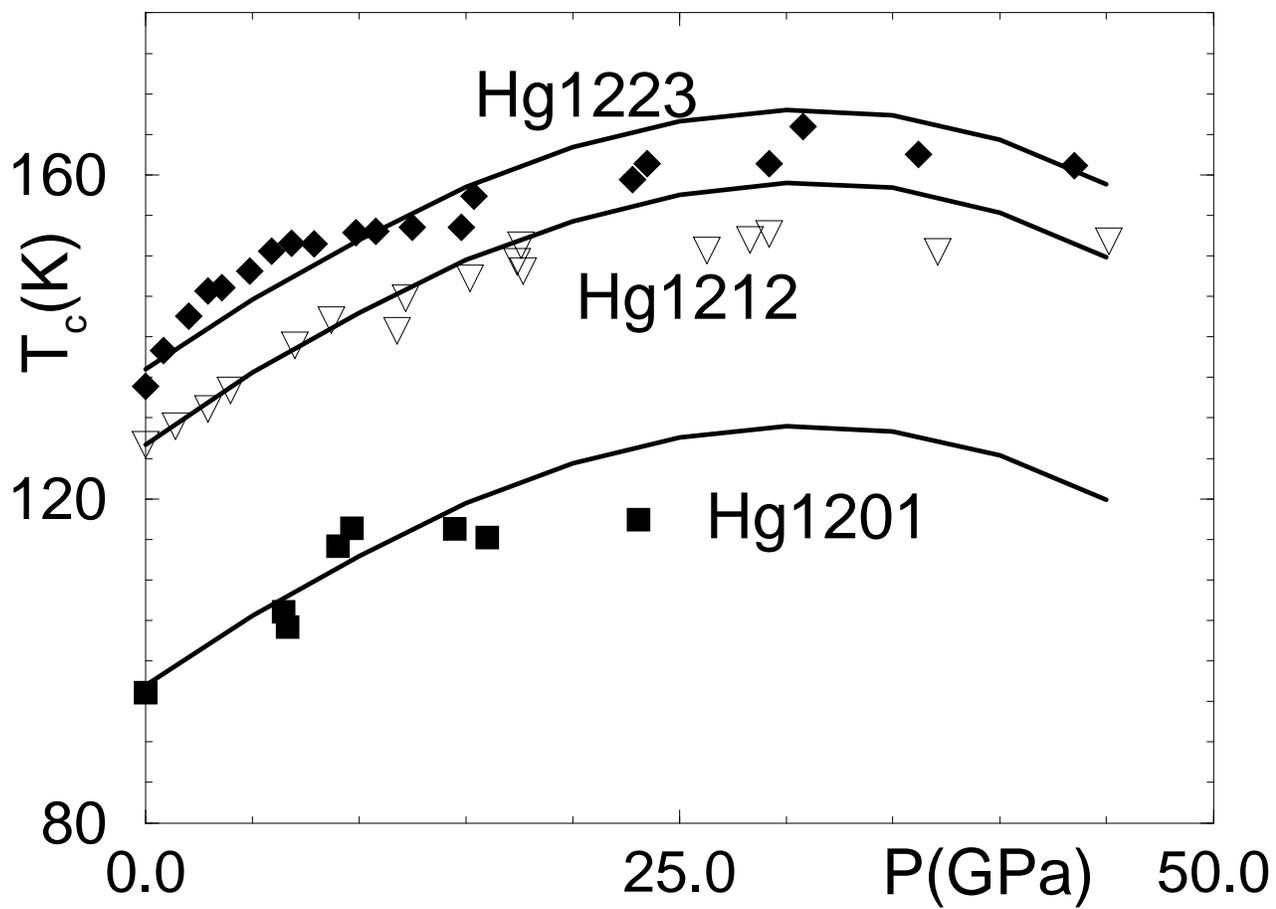
\begin{figure}
\caption{ The high pressures data of Ref.16  and our
calculations (the continuous lines) described in the text.}
\label{fig3}
\end{figure}

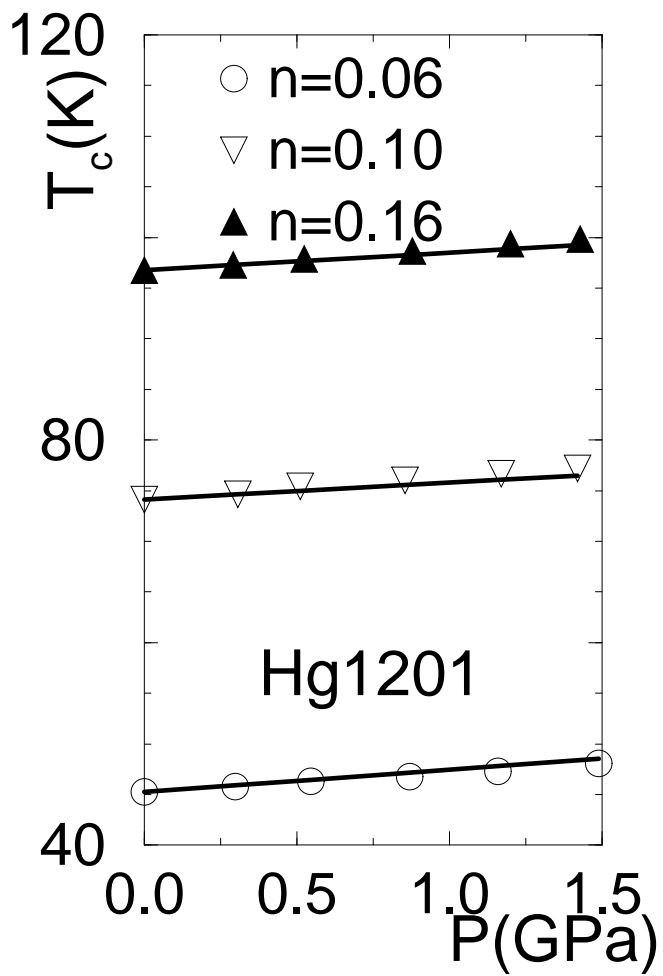
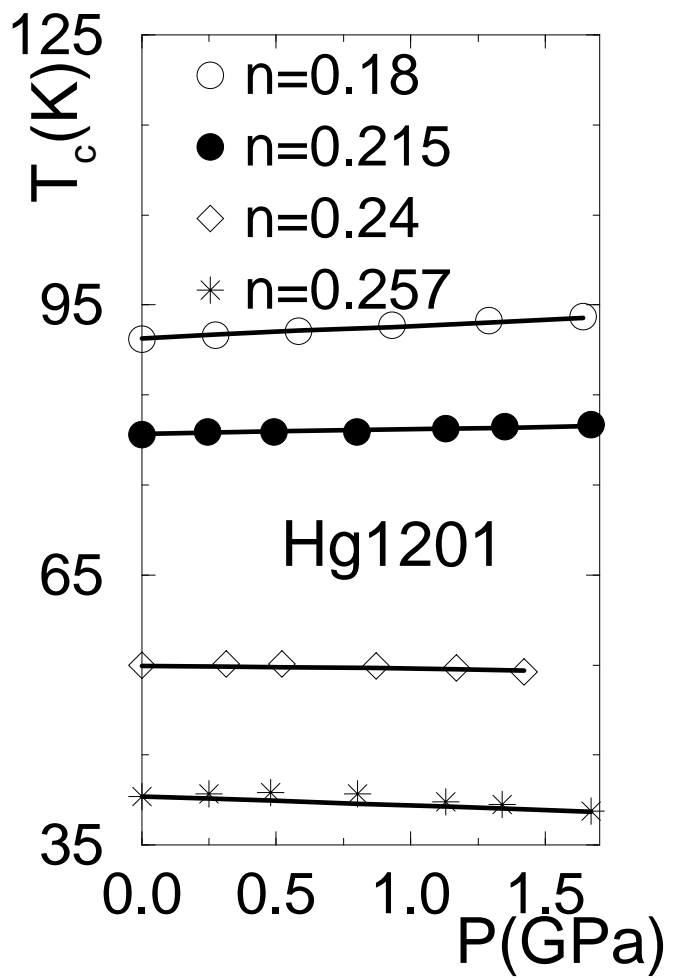
\end{document}